\documentclass[pra,aps,twocolumn,showpacs,superscriptaddress]{revtex4}
\usepackage{amssymb,amsmath,array,graphicx,epsfig,paralist}
\usepackage{subfigure}
\usepackage{here}

\def\fig_width{3.375 in} 
\def\fig_width{3. in} 
\draft

\newlength{\defbaselineskip}
\newcommand{\setlinespacing}[1]%
           {\setlength{\baselineskip}{#1 \defbaselineskip}}

\begin{document}

\title{Phase behavior of two-component lipid membranes: theory and experiments}

\author{Md. Arif Kamal}
\author{Antara Pal}
\author{V. A. Raghunathan}
\affiliation{Raman Research Institute, C V Raman Avenue, Bangalore 560 080, India}
\author{Madan Rao}
\affiliation{Raman Research Institute, C V Raman Avenue, Bangalore 560 080, India}
\affiliation{National Centre for Biological Sciences (TIFR), GKVK Campus, Bangalore 560 065, India}

\date{}


\begin{abstract}

The structure of the ripple phase of phospholipid membranes remains poorly understood in spite of a large number of theoretical studies, with many experimentally established structural features of this phase unaccounted for. In this article we present a phenomenological  theory of phase transitions in single- and two-component achiral lipid membranes in terms of two coupled order parameters -- a scalar order parameter describing {\it lipid chain melting}, and a vector order parameter describing the {\it tilt of the hydrocarbon chains} below the chain-melting transition.  This model reproduces all the salient structural features of the ripple phase, providing a unified description of the phase diagram and microstructure. In addition, it predicts a variant of this phase which does not seem to have been experimentally observed.  Using this model we have calculated generic phase diagrams of two-component membranes. We have also determined the phase diagram of a two-component lipid membrane from x-ray diffraction studies on aligned multilayers. This phase diagram is found to be in good agreement with that calculated from the model.

\end{abstract}

\maketitle
\section{Introduction}
Phospholipids are amphiphilic molecules with a hydrophilic head group and one or more hydrophobic hydrocarbon chains, and are the major constituents of biomembranes. They self-assemble in water to form a variety of thermodynamic phases, depending upon the temperature and water content~\cite{Tardieu1973}. The most common of which is the lamellar phase, a periodic structure consisting of lipid bilayers separated by layers of water.\par 
Above the chain melting transition (also known as the \emph{main-transition}) temperature $T_m$, phospholipids form the fluid $L_\alpha$ phase, where the  hydrocarbon chains are disordered with many \emph{trans-gauche} isomerizations in their C-C bonds. Positional order  in the plane of the bilayer is liquid-like in this phase. The lower temperature gel ($L_{\beta}$ or $L_{\beta^{\prime}}$) phase is characterized by flat bilayers with the hydrocarbon chains predominantly in the fully-stretched all-trans conformation. In the $L_{\beta^{\prime}}$ phase the chains are tilted relative to the bilayer normal, whereas in the $L_{\beta}$ phase there is no such tilt. The presence of a tilt depends on the relative cross-sectional areas of the head group and chains of the lipid species~\cite{MacIntosh1980,Miyata1996}. For example, phosphatidylcholines display a tilt angle of about 30$^\circ$ while  phosphatidylethanolamines with relatively smaller head groups do not show any tilt.\par
Phospholipids such as phosphatidylcholines (PC) exhibit a modulated phase between the $L_\alpha$ and the $L_{\beta^{\prime}}$ phases at high water content~\cite{Tardieu1973,Smith1988,Smith1990,Janiak1979}. This is known as the $P_{\beta^{\prime}}$ or the ripple phase, and  is characterized by a one-dimensional periodic height modulation of the bilayers. In this phase the hydrocarbon chains are partially ordered and have a non-zero tilt relative to the bilayer normal. The first order $L_{\beta^{\prime}}$ -- $L_{\alpha}$ transition is well understood in terms of the melting of some of the degrees of freedom within the hydrocarbon chains~\cite{Nagle1980}. However, the first-order $L_{\beta^{\prime}}$ --  $P_{\beta^{\prime}}$ transition (often called the \emph{pre-transition}), which occurs at a temperature slightly below the main transition, is much less understood.\par 
The structural features of the ripple phase have been extensively studied using a variety of techniques such as, x-ray diffraction (XRD)~\cite{Tardieu1973,Luna1977,Hansma1988,Hentschel1991,Wack1988}, neutron diffraction~\cite{Sackmann1988}, freeze-fracture electron microscopy~\cite{Luna1977,Sackmann1979,Hicks1987,Zasadzinski1988,Meyer1996}, and scanning-tunneling microscopy~\cite{Hansma1988,Woodward1997}. These experiments have revealed the existence of both \emph{symmetric} as well as \emph{asymmetric} ripples. The \emph{asymmetric }structure is, however, thermodynamically more stable and is commonly reported in almost all x-ray studies on PC systems. The wavelength of the ripples typically lies in the range of 12-16 nm~\cite{Sun1996} and the peak-to-peak amplitude is about 2 nm. X-ray diffraction patterns from this phase can be indexed on a two dimensional oblique lattice and the rippled bilayers lack a mirror plane normal to the rippling direction. Calculations of the electron density distributions~\cite{Wack1988,Sun1994,Kheya1999,Kheya2001,Kheya2000} confirm the basic  asymmetry of the ripples, and indicate that the bilayer thickness is appreciably different in the two arms of the ripple. The absence of a mirror plane normal to the ripple wave vector in the asymmetric ripple phase can be attributed to the presence of asymmetry in either the \emph{shape} (i.e., one arm being longer than the other) or \emph{bilayer thickness} (with a higher bilayer thickness in one of the arms) or both. However, in almost all the systems studied so far both these features are present simultaneously.\par
Several theoretical models have been proposed to describe the ripple phase~\cite{Doniach1979, Falkovitz1982, Marder1984, Sethna1987, Leibler1988, Leibler1989, Lubensky1993, Lubensky1995}, but none of them account for all the experimental observations, namely,\begin{inparaenum}[(i)]
\item occurrence of the $P_{\beta^{\prime}}$  phase between $L_\alpha$ and $L_{\beta^{\prime}}$ phases, separated by two first-order transitions;
\item unequal bilayer thickness in the two arms of the ripple; and
\item unequal lengths of the two arms.
\end{inparaenum}\par
Macroscopic continuum theories, which treat the bilayer as a continuous membrane, can broadly be classified into two main categories depending on the nature of the order parameter that is chosen to describe the ripple phase. One set of theories considers the order parameter to be a \emph{scalar}, such as the membrane thickness as in the models by Falkovitz \textit{et.al}~\cite{Falkovitz1982}, Marder \textit{et.al}~\cite{Marder1984} and by Goldstein and Leibler (GL)~\cite{Leibler1988,Leibler1989} or the configuration of the hydrocarbon chains (ratio of cis to trans bonds in the chains)~\cite{Kimura1991}. Due to the scalar nature of the order parameter, these models cannot distinguish the different symmetries of the $L_\alpha$, $L_{\beta^{\prime}}$ and $P_{\beta^{\prime}}$ phases. Further, in these models the ripple phase is characterized by a bilayer thickness modulation and not a height modulation. The other set of continuum theories considers the order parameter to be a \emph{vector}, such as the tilt of the hydrocarbon chains with respect to the local layer normal, as in the Lubensky-MacKintosh (LM)~\cite{Lubensky1993} and the Seifert-Shillcock-Nelson models~\cite{Seifert1996}. \par
Although models involving vectorial order parameter can predict the various phases with different in-plane symmetries, they have their own limitations. For example, the LM-model cannot account for the occurrence of the ripple phase in bilayers composed of achiral molecules and speculated that the the origins of the experimentally observed asymmetric ripples lay in the chirality of lipid molecules. However, subsequent experiments using racemic (achiral) mixtures showed this was not the case ~\cite{Katsaras1997,Raghu1995}.\par
In this paper we present a phenomenological Landau model to describe the ripple phase, that unifies the two distinct approaches (those using scalar and vector order parameters) used in the earlier models. In addition to the coupling between the curvature of the bilayer to the divergence of the molecular tilt~\cite{Lubensky1993}, we bring out the importance of the coupling of  variation in the bilayer thickness to molecular tilt and show that the
\begin{inparaenum}[(i)]
\item asymmetry of the ripple phase arises from 
						\begin{inparaenum}[(a)]
						\item the difference in the bilayer thickness in the two arms of the ripple,
						\item unequal length of the two arms, 
						\end{inparaenum}
\item existence of a nonzero mean tilt of the hydrocarbon chains of the lipid molecules along the direction of ripple.
\end{inparaenum}
In section II we present the model for single-component achiral membranes. The main results of this section has already been reported in a short communication \cite{Kamal2011}. In section III we extend the model to two-component membranes and use it to calculate the phase diagrams for various binary lipid mixtures. In section IV we present the phase diagram of dipalmitoyl phosphatidylcholine (DPPC) - dimyristoyl phosphatidylcholine (DMPC) mixtures determined from small-angle x-ray scattering studies of aligned samples. The experimental phase diagram is found to be in good agreement with the one calculated using our model.
\section{Single-component achiral membranes}
\begin{figure}[htbp]
	\begin{center}
		\includegraphics[scale=0.90] {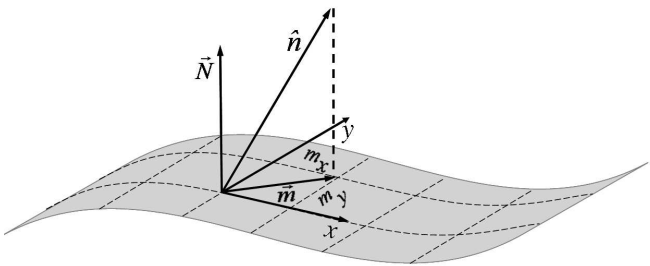}
	\end{center}
	\caption{The unit vector $\hat{\bf{n}}$ represents the orientation of the long axis of the lipid molecules relative to ${\bf{N}}$, the bilayer normal. The projection of $\hat{\bf{n}}$ on the bilayer plane ${\bf{m}}\:(= \hat{\bf{n}} - ({\bf{N}}\cdot \hat{\bf{n}})\:\hat{\bf{N}})$ is the order parameter. $m_x$ and $m_y$ are the components of ${\bf{m}}$ along the two orthogonal directions in the plane of the bilayer.}
\label{tlt_m}
\end{figure}

In this section, we develop a phenomenological model for describing the asymmetric ripple phase in achiral membranes. To simplify matters, we ignore interbilayer interactions and choose to describe the ripple phase in terms of
\begin{inparaenum}[(i)]
	\item a scalar order parameter $\psi$, which represents the relative bilayer thickness and 
	\item a vector order parameter \textbf{m}, which is the projection of the molecular axis in the plane tangent to the bilayer surface (fig.(\ref{tlt_m}))~\cite{Lubensky1993}.
\end{inparaenum}	
We consider three distinct contributions to the total free energy density:  
\begin{description} \item [a.]\textsf{The stretch free energy density $f_{st}$ :}\end{description}\par
The main transition is characterized by the freezing of the lateral motion as well as conformational ordering of the hydrocarbon chains. Many experiments have revealed that at the main transition there is a discontinuous jump in the membrane thickness~\cite{Rand1981}. Therefore, following GL, we describe the main transition as  one involving a change in the bilayer thickness and introduce a dimensionless scalar order parameter $\psi$, such that:
\begin{equation}
\psi\:=\:\frac{\delta(\textbf{r})\:-\:\delta_0}{\delta_0}
\end{equation}
where $\delta(\textbf{r})$ is the actual membrane thickness, $\delta_0$ is the constant thickness of the membrane in the $L_\alpha$ phase and \textbf{r} denotes the lateral position within the bilayer plane. No specific assumption is made about the microscopic state of the lipids in choosing $\psi$ as the order parameter; we can think of $\psi$ as a hybrid coarse grained variable that summarizes changes in various types of degrees of freedom, like for example, conformational changes of the hydrocarbon chains. $\psi$ is taken to be positive for $T$ $<$ 
$T_m$ due to the stretching of the chains. This is valid in general, even if the chains are tilted below $T_m$.\par

The stretching free energy per unit area, for an isolated lipid bilayer membrane, can therefore be written as~\cite{Komura2008};
\begin{align}
f_{st}=&\frac{1}{2}\:a_2\:\psi^2\:+\:\frac{1}{3}\:a_3\:\psi^3\:+\:\frac{1}{4}\:a_4\:\psi^4\:
+\:\frac{1}{2}C\left(\nabla\psi\right)^2\nonumber
\\ &+\:\frac{1}{2}D\left(\nabla^2\psi\right)^2+\:\frac{1}{4}E\:\left(\nabla\psi\right)^4 
\label{F_st}
\end{align}
The explicit temperature dependence is assumed to reside solely in the coefficient of $\:\psi^2$ : 
	\[a_2\:=\:a_2^{\prime}\:(T-T^*)
\]
$T^{\ast}$ being a reference temperature (which in the absence of the cubic term in $f_{st}$ is the critical temperature). Since, $L_{\beta^{\prime}}\to L_{\alpha}$ phase transition is known to be first order, the coefficient $a_3$ is taken to be negative. In the presence of the cubic term, the critical temperature $T^{\ast}$ is preempted by a first order melting transition at,
\begin{equation}
T_m\;=\;T^*\;+\;\frac{2 a_3^{2}}{9a_2^{\prime} a_4}
\label{Ref_Temp}
\end{equation}
The coefficient C can either be positive or negative but D , E and $a_4$ are always positive to ensure the stability of this free energy expansion. With $C>0$, the equilibrium phases are always spatially homogeneous; either as $L_\alpha$ or 
$L_{\beta}$ ($L_{\beta^{'}})$.
However, with $C<0$ modulated phases are possible with some characteristic wave vector $q_0$.   To motivate how $C$ can be negative, it is convenient to introduce an auxiliary scalar variable $\rho$. Constructing  effective cylinders enclosing the head group and the chain with cross-sectional areas $a_h$ and $a_t$, respectively,  we define $\rho \equiv (a_h -a_t) /(a_h+a_t)$.
Consider a bilayer membrane whose thickness $\psi$ increases steadily over a length scale $\xi$. This variation in thickness produces a strain at the molecular level. This can be accommodated both by a variation in the tilt $\bf m$,
 and a variation 
in the mismatch $\rho$ over this scale. The latter coupling to lowest order is of the form 
$\nabla \psi \cdot \nabla \rho$. This term,
consistent with the symmetries of the bilayer, 
will generically lead to $C<0$.  

\begin{description} \item [b.]\textsf{The tilt free energy density $f_{tilt}$ :}\end{description}\par 
The tilt free energy density  can then be written in terms of ${\bf{m}}$ as,
\begin{align}
f_{tilt}=& \frac{1}{2}b_{2}\left|\textbf{m}\right|^2+ \frac{1}{4}b_{4}\left|\textbf{m}\right|^4+
{\Gamma}_1\left(\nabla\cdot\textbf{m}\right)^2 \nonumber\\
&+\Gamma_2\left(\nabla^2\textbf{m}\right)^2 +\Gamma_3\left(\nabla\cdot\textbf{m}\right)^4
+\Gamma_{4}\psi\left|\textbf{m}\right|^2 \nonumber\\
&+\Gamma_{5}\left(\textbf{m}\cdot\nabla\psi\right)^2+\Gamma_{6}\left(\textbf{m}\times\nabla\psi\right)^2\nonumber\\
&+\Gamma_{7}\left(\nabla^{2}\psi\right)\left(\nabla\cdot\textbf{m}\right)^2
+\Gamma_{8}\left(\nabla\psi\right)^2\left(\nabla\cdot\textbf{m}\right)^2
\label{F_tlt}
\end{align}
First and second terms in the above equation represent the coupling of the molecular tilt vector \textbf{m} to the bilayer thickness and  modulations in the thickness, respectively. The third term corresponds to the splay deformation of the \textbf{m} field and represent the tilt elasticity in the tilted phase. Since the order parameter {\bf {m}} is a vector, cubic terms in {\bf {m}} are not allowed.\par
As shown by Jacobs \textit{et.al}~\cite{Jacobs1984} in the context of a one dimensional model with a scalar order parameter, a term in the fourth power of the gradient of the order parameter is necessary to stabilize a modulated phase with a non zero mean value of the order parameter. Hence we include $\left(\nabla\psi\right)^4$ and $\left(\nabla\cdot\textbf{m}\right)^4$ terms in our expression for stretching free energy density and the tilt free energy density, respectively.\par
As $b_2$ is positive in our model, the term involving the coefficient $\Gamma_4$, which couples the molecular tilt to the change in the bilayer thickness, is the most crucial one for the appearance of the asymmetric ripple phase. When $\Gamma_4$ $>$ 0, the stable phase below $T_m$ is $L_\beta$ with $\left|\textbf{m}\right|$=0. On the other hand, tilted phases can form if $\Gamma_4$ $<$ 0. The succeeding two terms take into account the anisotropy of the tilted bilayer. The next two terms represent higher order couplings between modulations in $\psi$ and in $\bf{m}$, allowed by the symmetry of the system. These higher order coupling terms are kept merely for consistency and are not crucial for obtaining the optimal phases.
\begin{description} \item [c.]\textsf{The curvature energy density $f_{c}$ :}\end{description}\par 
The curvature energy density of the bilayer can be written as ~\cite{Lubensky1993};
\begin{equation}
f_c\:=\:\frac{1}{2}\:\kappa\:\left(\nabla^2\:h\right)^2\:-\:\gamma\left(\nabla^2\:h\right)\:\left(\nabla\cdot\textbf{m}\right)
\label{F_curv}
\end{equation}
where $h(x,y)$ is the height of the bilayer relative to a flat reference plane, $\kappa$ is the bending rigidity of the membrane, and the coefficient $\gamma$ couples the mean curvature to splay in \textbf{m}.\\ 
The total free energy density is given by the sum of the above three contributions :
\begin{equation}
f= f_{st}\:+\:f_{tilt}\:+\:f_c
\end{equation}
The equilibrium height profile of the bilayer $h(x,y)$ is related to the tilt \textbf{m} via the Euler-Lagrange Equation,
\begin{equation}
\nabla^2\:h\:=\:\frac{\gamma}{\kappa}\:\left(\nabla\cdot\textbf{m}\right)
\end{equation}
Eliminating $h$ from the free energy density $f$ leads to the effective energy density $f_{eff}$ with a reduced splay elastic constant.
\begin{equation*}
\tilde\Gamma_1= {\Gamma_1}-\frac{\gamma^2}{\kappa}\nonumber
\end{equation*}
Thus, the effective free energy density is given by,
\begin{eqnarray}
f_{eff} &=&\frac{1}{2}\:a_2\:\psi^2\:+\:\frac{1}{3}\:a_3\:\psi^3\:+\:\frac{1}{4}\:a_4\:\psi^4\:+\:\frac{1}{2}\:C\left(\nabla\psi\right)^2\nonumber\\
				& & +\:\frac{1}{2}\:D\left(\nabla^2\psi\right)^2+\:\frac{1}{4}E\:\left(\nabla\psi\right)^4 +\:\frac{1}{2}\:b_{2}\left|\textbf{m}\right|^2\nonumber\\
				& & +\:\frac{1}{4}\:b_{4}\left|\textbf{m}\right|^4\:+\:\tilde{\Gamma}_1\left(\nabla\cdot\textbf{m}\right)^2\:+\:\Gamma_2\left(\nabla^2\textbf{m}\right)^2\nonumber\\
				& & +\:\Gamma_3\left(\nabla\cdot\textbf{m}\right)^4\:+\:\Gamma_{4}\psi\left|\textbf{m}\right|^2\:+\:\Gamma_{5}\left(\textbf{m}\cdot\nabla\psi\right)^2\nonumber\\
				& & +\:\Gamma_{6}\left(\textbf{m}\times\nabla\psi\right)^2\:+\:\Gamma_{7}\left(\nabla^{2}\psi\right)\left(\nabla\cdot\textbf{m}\right)^2\nonumber\\
				& & +\:\Gamma_{8}\left(\nabla\psi\right)^2\left(\nabla\cdot\textbf{m}\right)^2
				\label{F_eff}
\end{eqnarray}
\par
To determine the mean field phase diagram we assume that the modulated phase is characterized by the most dominant wave vector q, and choose the following ansatz for $\psi$ and $\textbf{m}$,
\begin{align}
\psi\ &=\:\psi_0+\:\psi_1sin(qx) \nonumber \\
m_x&=\:m_{0x}+\:m_{1x}\:cos\left(qx\right)+m_{2x}\:sin\left(qx\right)\nonumber\\ 
&\phantom{{=}m_{0x}{+}}+\:m_{3x}\:cos\left(2qx\right)+m_{4x}\:sin\left(2qx\right)\nonumber \\
m_y&=\:m_{0y}
\label{ansatz}
\end{align}   
with the amplitudes $\left\{{\psi_i},m_i\right\}$ as variational parameters. We retain Fourier components of $m_x$ to second order to account for the ripple asymmetry; the experimentally observed asymmetric ripple profile is obtained if $ m_{3x} \neq 0 $. In principle, one could have done a numerical variational calculation with many more Fourier components. However, our simplified ansatz is sufficient to recover all the qualitative features of the experimentally observed asymmetric ripple phase. \par
2D modulations of the membrane~\cite{Yang2005} was not included in our ansatz (eqn.[\ref{ansatz}]), since earlier studies~\cite{Lubensky1993,Lubensky1995,Kheya2001} based on a similar free energy density for the tilt and curvature alone, show that the presence of a mean tilt suppresses two dimensional height modulations. The additional $\psi$ dependent terms in the free energy density do not alter this conclusion. Hence we confine our attention to one-dimensionally modulated ripples. Note that spatial modulations in $m_y$ are neglected as we do not keep terms proportional to $(\nabla \times m)$ in eqn.(\ref{F_tlt}). Variation of $m_y$ with $x$ represents non-parallel orientations of the lipid chains, which are not favored by van der Waal's interactions between the chains. Such terms are important in chiral systems, where they can lead to structures with non-zero winding numbers~\cite{Lubensky1993}. In the absence of chirality, these terms can, however, be neglected. \par
The average free energy is obtained by integrating the $F_{eff}$ over one spatial period,i.e,
\begin{equation}
\left\langle f_{eff}\right\rangle\:=\:\frac{\int_{0}^{\frac{2 \pi}{q}} f_{eff}\:dx}{\int_{0}^{\frac{2 \pi}{q}}\:dx}
\end{equation}
\subsection{Results and discussion}

\begin{figure}[htbp]
	\begin{center}
		\includegraphics[scale=0.55] {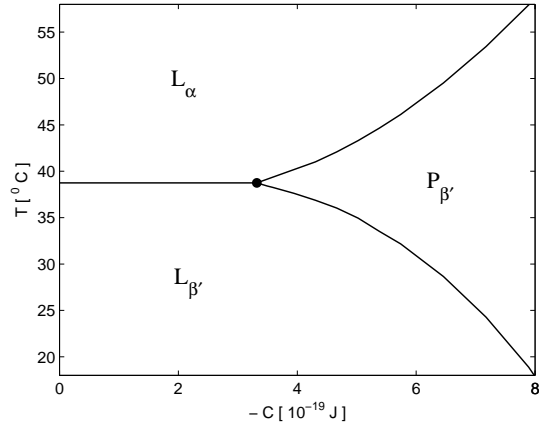}
	\end{center}
	\caption{Phase diagram in the $T-C$ plane calculated from the model. $a_2^\prime$ = 159.42 $k_B$, $T^*$ = 260.0 K. Values of the  other coefficients in units of $k_BT^*$ are: $a_3$ = -306.5, $a_4$ = 613.15, $b_2$ = 0.2, $b_4$ = 200.0, $D$ = 557.41, $E$ = 600.0, $\Gamma_1$ = 0.010, $\Gamma_2$ = 1.80, $\Gamma_3$ = 500.0, $\Gamma_4$ = -3.0, $\Gamma_5$ = -20.0, $\Gamma_6$ = -20.0, $\Gamma_7$ = -50.0, $\Gamma_8$ = -25.0. Both the main-transition ($L_\alpha$ $\rightarrow$ $L_{\beta^\prime}$; $L_\alpha$ $\rightarrow$ $P_{\beta^\prime}$) and pre-transition ($P_{\beta^\prime}$ $\rightarrow$ $L_{\beta^\prime}$) are first order.} 
\label{phasedia}
\end{figure}

The phase diagram in the C-T plane is obtained from the numerical minimization of the effective free energy $\left\langle f_{eff}\right\rangle$. Fig.(\ref{phasedia}) represents an example of such a phase diagram. 
It is calculated for a choice of parameter values, which reproduce closely the main- and pre-transition temperatures of  DPPC  at $C$ = - 4.84$\times10^{-19}J$. These values (except for $\Gamma_1$ to $\Gamma_6$) are taken from ref~\cite{Leibler1989} and ref~\cite{Komura2008}. It is seen from fig.(\ref{phasedia}) that there are three distinct regions in the phase diagram corresponding to three different phases:
\begin{equation*}
L_{\alpha}\:\text{phase\::}
\begin{cases}
\psi_0=0,\ \psi_1=0,\ m_{0x}=0,\ m_{0y}=0 \\
m_{1x}=0,\ m_{2x}=0,\ m_{3x}=0,\ m_{4x}=0
\end{cases}
\end{equation*}
\begin{equation*}
L_{\beta^{'}}\:\text{phase\::}
\begin{cases}
\psi_0\neq0,\  \psi_1=0,\  m_{0x}\neq0,\ m_{0y}\neq0 &\\
m_{1x}=0,\ m_{2x}=0,\ m_{3x}=0,\ m_{4x}=0
\end{cases}
\end{equation*}
\begin{equation*}
P_{\beta^{'}}\:\text{phase\::}
\begin{cases}
\psi_0\neq0,\  \psi_1\neq0,\  m_{0x}\neq0,\ m_{0y}=0 &\\
m_{1x}=0,\ m_{2x}\neq0,\ m_{3x}\neq0,\ m_{4x}=0
\end{cases}
\end{equation*}
\par
The first-order transition lines which separate these three phases meet at a Lifshitz point at $C_{Lp}= - 3.43 \times10^{-19}J$ and $T = 38.75^{\circ}C$. When $\left|C\right|<\left|C_{Lp}\right|$, the first order transition line is parallel to the C axis and occurs at $T=38.75^{\circ}C$. As long as $\left|C\right|<\left|C_{Lp}\right|$ the $L_{\beta^{\prime}}$ phase melts directly to the $L_{\alpha}$ phase at $T_m$. But for $\left|C\right|>\left|C_{Lp}\right|$ the $L_{\beta^{'}}$ phase first melts into the $P_{\beta^{\prime}}$ phase, which in turn transforms into the $L_{\alpha}$ phase at higher temperatures. It is also clear from the figure that, as $\left|C\right|$ increases, the region of the $P_{\beta^{\prime}}$ phase expands at the expense of the $L_{\beta^{\prime}}$ and the $L_{\alpha}$ phases. \par
\begin{figure}[htbp]
	\begin{center}
		\includegraphics[scale=0.4] {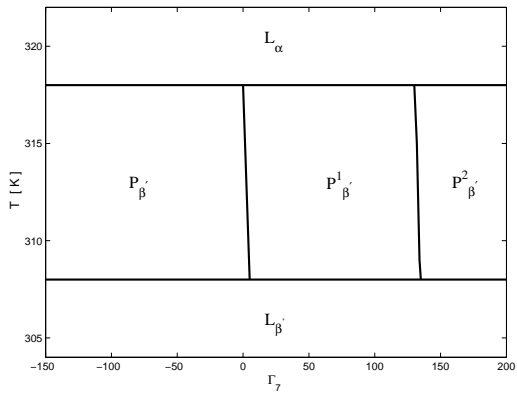}
	\end{center}
	\caption{Dependence of the phase behavior on $\Gamma_7$. All lines in the phase diagram correspond to first-order transitions. C = 5.74$\times$10$^{-19}$ J, and all other parameters as in fig.\ref{phasedia}.} 
\label{g7-t}
\end{figure}
\begin{figure}[h!]
\centering
\subfigure[]{
\includegraphics[scale=0.4]{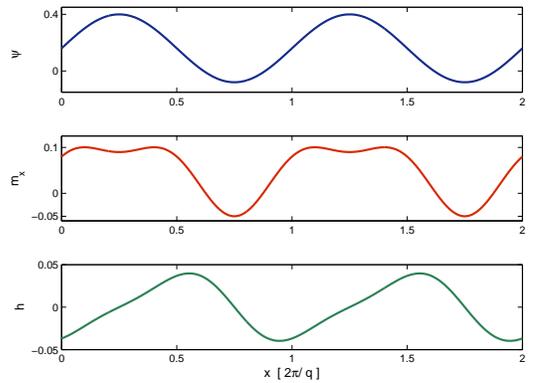}
\label{fig:opsg7-150}
}
\subfigure[]{
\includegraphics[scale=0.425]{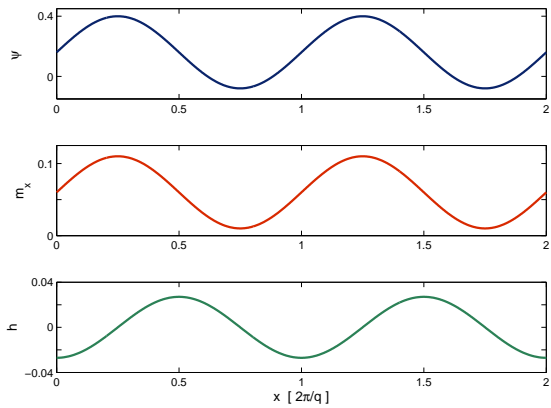}
\label{fig:opsg7100}
}
\subfigure[]{
\includegraphics[scale=0.425]{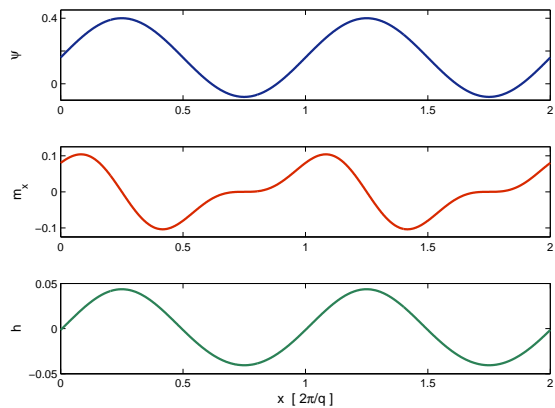}
\label{fig:opsg7150}
}
\caption{Spatial variation of $\psi$, $m_x$ and $h$ in the different ripple phases obtained for \subref{fig:opsg7-150} $\Gamma_7 = -150\:k_BT^{\ast} $, \subref{fig:opsg7100} $\Gamma_7 =100\:k_BT^{\ast}$ and \subref{fig:opsg7150} $\Gamma_7 = +150\:k_BT^{\ast}$. T = 310.50 K and all other parameters are the same as those used in fig.\ref{phasedia}.}
\end{figure}

 It would seem that our free energy functional contains a large number of parameters, which we need to fine tune to obtain the asymmetric ripple phase. We would like to emphasize that asymmetric ripples are a fairly generic and robust feature of our model {\it whenever  $\Gamma_4 <0$}. The other $\Gamma_i$'s  only make  the shape anisotropy in the height profile more pronounced when they are negative, and are not crucial for the existence of the asymmetric ripple phase. Further, x-ray diffraction experiments identify
the asymmetric ripple phase on the basis of its oblique unit cell. Such a unit cell can arise
either from a strongly asymmetric height profile or from a weakly asymmetric height profile
as long as the thickness is different in the two arms of the ripple \cite{Kheya2001}. This feature is exactly what is 
captured within our theory (Figs. \ref{fig:opsg7-150} and \ref{fig:opsg7100}).

In addition to $C$ and $\Gamma_4$, the phase behavior depends crucially on $\Gamma_7$. Depending on the value of $\Gamma_7$ three distinct modulated phases are found, with differing spatial variations of the order parameters (fig.\ref{g7-t}). For $\Gamma_7$ $<$ 0, the $P_{\beta^\prime}$ phase is stable, where the height profile is asymmetric and resembles very closely those seen in experiments (fig.\ref{fig:opsg7-150})~\cite{Sun1996,Kheya2003,Woodward1997}.  In this phase $m_{1x}=0,\ m_{2x}\neq0,\ m_{3x}\neq0,\ m_{4x}=0$. $\psi_1$ is almost $\pi/2$ out of phase with h, so that it is positive (negative) along the longer (shorter) arm of the ripple, resulting in different bilayer thicknesses in the two arms, again in agreement with experimental observations~\cite{Sun1996,Kheya2003}. When $\Gamma_7$ is weakly positive, the $P^1_{\beta^\prime}$ is obtained (fig.\ref{fig:opsg7100}), where the higher Fourier components of $m_x$ are absent ($m_{1x}=0,\ m_{2x}\neq0,\ m_{3x} = 0,\ m_{4x}=0$) and hence the height profile is symmetric. As in the $P_{\beta^\prime}$ phase the bilayer thickness is different in the two arms. This profile resembles the structure reported in ref. \cite{Kheya2001}, based on the diffraction data of ref. \cite{Tardieu1973}. For higher values of $\Gamma_7$ 
the ${P^2}_{\beta^\prime}$ is found (fig.\ref{fig:opsg7150}), where $m_{1x}\neq 0,\ m_{2x} = 0,\ m_{3x}= 0,\ m_{4x} \neq 0$. The height profile is symmetric and is almost in phase with $\psi_1$. As a result, the bilayer thickness is modulated within each arm of the ripple. This novel structure does not seem to have been observed in any experiments till now. However, since the electron density map of the ripple phase has been calculated only in a few instances, the possible occurrence of this structure in some systems cannot be ruled out.

The present model predicts the chain tilt to be high in the thicker arm and negligible in the thinner one. There are in principle at least two chain conformations consistent with this thinner arm of the bilayer -- one is by having disordered chains, the other by chain interdigitation. Since in our model, the high temperature phase was identified as $L_{\alpha}$, it is natural to populate the
thinner arm with disordered chains. We therefore propose 
 that in the asymmetric ripple,  the thicker arm is  made up of tilted gel-like domains and the thinner one is fluid-like with negligible chain tilt. This conclusion is partially supported by
 spectroscopy and diffusion experiments, which indicate a significant fraction of disordered chains in the ripple phase \cite{Wittebort81,Schneider83}. However, it must be acknowledged that there is  no direct experimental data on the spatial distribution of the disordered chains within the rippled bilayer. Electron density maps calculated from diffraction data do not provide any direct information about chain ordering. 
In contrast, recent computer simulation studies of the ripple phase using a variety of model lipids at different levels of coarse-graining \cite{deVries05,Lenz07,Jamroz10}, find 
 an asymmetric ripple structure consistent with the height and thickness modulations seen in the electron density maps,
with   the thicker arm consisting of gel-like domains, and the thinner one made up of ordered and interdigitated chains. The cross-over region between these two contains a large fraction of disordered chains. In the gel-like domains chains are tilted with respect to the local layer normal, whereas chain tilt vanishes in the thinner arm due to interdigitation. Thus the height, thickness and tilt modulations found in these simulations  are consistent with the predictions of the present model, although the thinner arm consists of interdigitated chains instead disordered chains as proposed here. Symmetric ripples consisting of alternating gel-like and fluid-like domains have also been found in some simulations \cite{Kranenburg05}. These ripples are characterized by only a thickness modulation and no height modulation. Stacks of such bilayers will result in a centered rectangular lattice and do not seem to have been observed in any pure lipid-water system.

\section{Binary Lipid mixtures}
{Biomembranes are multicomponent, including lipid species that show liquid-ordered behavior, and are likely to have thickness variations \cite{simons11}. The first step in understanding such complex systems is to study binary mixtures of lipids which differ in their head groups and chains. Therefore, in this section, we  extend the model presented in the previous section for single component membranes to membranes consisting of two lipid species. }

We specifically consider two different cases. In case 1, only one of the lipids exhibits the ripple phase. Since the occurrence of the ripple phase is closely related to the formation of a tilted gel phase at lower temperatures,  one of the lipids is assumed to show the phase sequence $L_{\alpha}$ $\rightarrow$  $P_{\beta^{\prime}}$ $\rightarrow$ $L_{\beta^{\prime}}$ , and the other to show $L_{\alpha}$ $\rightarrow$ $L_{\beta}$ on cooling. This case corresponds to two species differing only in their head groups.  In case 2, both the lipids exhibit the ripple phase, and corresponds to the situation where the two lipids differ only in the length of their hydrocarbon chains. We employ an approach quite similar to that used in ref~\cite{Komura2008} and ignore any lipid exchange with the surrounding solvent. Owing to the fact that the lipids constituting the mixture can have different chain lengths, degrees of saturation, or head groups, the  main transition temperature is taken to be different for the two lipids.\par
Due to the presence of an additional component in the system there is a contribution $f_{mix}$ to the total free energy, corresponding to the free energy of mixing. $f_{mix}$ is given by the sum of the entropy of mixing and enthalpy, which within the Bragg-Williams (mean-field) approximation can be written as,
\begin{equation}
f_{mix}=k_B T\left[\phi\:log(\phi)+(1-\phi)\:log\:(1-\phi)\right]+\frac{1}{2}J\phi\:(1-\phi)
\label{F_mix}
\end{equation}
where, $k_B$ is the Boltzmann constant, and $J>0$ is an attractive interaction parameter that enhances lipid-lipid demixing. \par
Since the important properties of the lipids are mainly reflected in the model parameters $T^{\ast}$, C and $\Gamma_4$, we assume their values for a mixture to be the average of that for the two pure lipids weighted by their concentration. 
\begin{eqnarray}
	C(\phi)&=&\:\phi\:C_A\ +(1-\phi)\:C_B\\
	\Gamma_4(\phi)&=&\:\phi\:\Gamma_{4A}+(1-\phi)\:\Gamma_{4B}\\	
	T^{\ast}(\phi)&=&\:\phi\:T_A^{\ast}+(1-\phi)\:T_B^{\ast}
	\label{para-mix}
\end{eqnarray}
where, the subscripts A and B refer to the lipids A and B, respectively, and $\phi$ is the mole fraction of lipid A in the mixture. All other parameters are assumed to be same for both the lipids.\\
Combining eqn.(\ref{F_eff}) and eqn.(\ref{F_mix}), we obtain the total free energy density $f_{total}$ as,
\begin{equation}
f_{total}=\:f_{mix}+\:f_{eff}
\label{F_total}
\end{equation}
After minimizing Eqn.(\ref{F_total}) with respect to both $\psi$ and $\textbf{m}$, the two-phase coexistence in the $(T,\phi)$ plane is obtained using Maxwell's construction.
\subsection{Results and discussion}
\subsubsection{ Tilt-Non-Tilt Mixtures}
We now calculate the phase diagram of binary lipid mixtures, where only one of the constituents exhibits the rippled phase, such as mixtures of dipalmitoyl phosphatidylcholine (DPPC) and dipalmitoyl phosphatidylethanolamine (DPPE). The phosphatidylcholines, which have the $-N(CH_3)_3^{+}$ moiety in their polar head groups, typically exhibit a tilted $L_{\beta^{\prime}}$ phase. On the other hand, the phosphatidylethanolamines which have the smaller $-(NH_3)_3^{+}$ moiety instead, shows the untilted $L_{\beta}$ phase and do not exhibit the ripple phase.\par
From fig(\ref{phasedia}) one finds that only when the value of $\left|C\right|$ is greater than $\left|C_{Lp}\right|$, a spatially modulated phase occurs between the high and low temperature phases. The sign of $\Gamma_{4}$ is also important  in determining the existence of a tilted phase at lower temperatures. If $\Gamma_{4} < 0 $, then the tilted phase $L_{\beta^{\prime}}$ occurs, otherwise one has the non tilted $L_{\beta}$ phase below the main transition temperature.\par
Keeping this in mind, we choose the value of $\left|C\right|$ for lipid A (DPPE) to be less than $\left|C_{Lp}\right|$ and that for lipid B (DPPC) to be greater than $\left|C_{Lp}\right|$. The parameter values are chosen so as to reproduce the transition temperatures of DPPC and DPPE. $\Gamma_4$ is chosen to be negative for DPPC and positive for DPPE. \par
\begin{figure}[t!]
	\centering
		\includegraphics[scale=0.55]{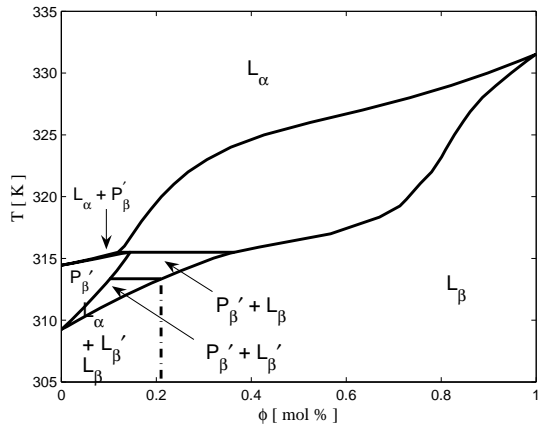}
	\caption{\footnotesize{Calculated mean-field phase diagram of DPPC- DPPE mixture as a function of composition and temperature T. $\phi$ is the mole fraction of DPPE. C$_{DPPC}$ = 4.66$\times$10$^{-19}$J, C$_{DPPE}$ = 3.58$\times$10$^{-21}$ J, ${T^{\ast}}_{DPPC}$ = -17$^{\circ}$C, ${T^{\ast}}_{DPPE}$= -3$^{\circ}$C, and $J = 4.0\:k_BT^{\ast}$. All other parameter values are as in fig.\ref{phasedia}.}}
	\label{fig:TNT_J4}
\end{figure}
\begin{figure}[t!]
	\centering
	\includegraphics[scale=0.55]{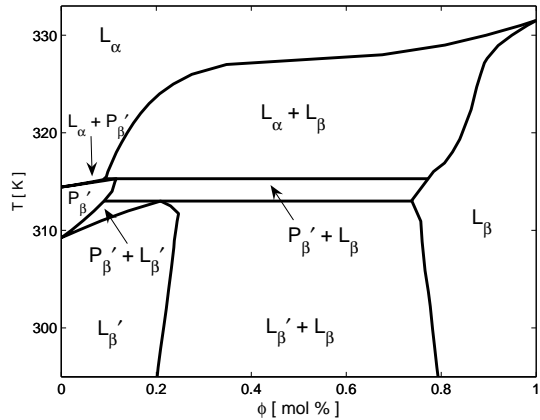}
	\caption{\footnotesize{Calculated mean-field phase diagram of a binary lipid mixture as a function of composition and temperature at J = 5.0$k_BT^{\ast}$. All other parameter values are as in fig.\ref{fig:TNT_J4}.}}
		\label{fig:TNT_J5}
\end{figure}
Using the above values of the parameters the binary phase diagram for DPPC-DPPE mixtures were calculated, and the results are shown in fig.\ref{fig:TNT_J4} and fig.\ref{fig:TNT_J5}. In fig.\ref{fig:TNT_J4} the interaction parameter $J = 4.0\:k_{B} T^{*}$ while in fig.\ref{fig:TNT_J5} $J = 5.0\:k_{B} T^{*}$. \par
The calculated phase diagram of fig.\ref{fig:TNT_J4} is found to be in good quantitative agreement with the experimental one for DPPC/DPPE mixtures~\cite{Blume1982}, and is very similar to what is obtained for other lipid mixtures such as, DMPC/ dipalmitoyl phosphatidylserine (DPPS) and DPPC/DPPS ~\cite{Luna1977}. At higher temperatures,  we have the $L_{\alpha}$ phase for the entire range of composition. With decreasing temperature the $P_{\beta^{\prime}}$  phase appears very close to the pure DPPC axis, and is bounded by coexistence regions with the $L_{\beta^{\prime}}$, $L_\beta$ and the $L_{\alpha}$ phases. The $P_{\beta^{\prime}}$ - $L_{\alpha}$ coexistence region is very narrow on the scale of the figure and collapses almost into a line. At low temperatures there is a continuous $L_{\beta^{\prime}}$ - $L_{\beta}$ transition at $\phi \sim 0.2$, which reflects the change in the sign of $\Gamma_4$ according to eqn.\ref{para-mix}.\par
Fig. \ref{fig:TNT_J5} shows the phase diagram for a binary lipid system in which the constituent lipids show a very high tendency to segregate and resembles the experimentally obtained one for dielaidoyl phosphatidylcholine (DEPC)/DPPE mixtures ~\cite{Wu1975}. The two lipids are completely miscible only in the $L_{\alpha}$ phase. The main qualitative difference from the previous case is the large region of  $L_{\beta^{\prime}}$ - $L_{\beta}$ coexistence at lower temperatures.
\subsubsection{ Tilt-Tilt Mixtures}
\begin{figure}[h!]
\centering
\subfigure[]{
\includegraphics[scale=0.55]{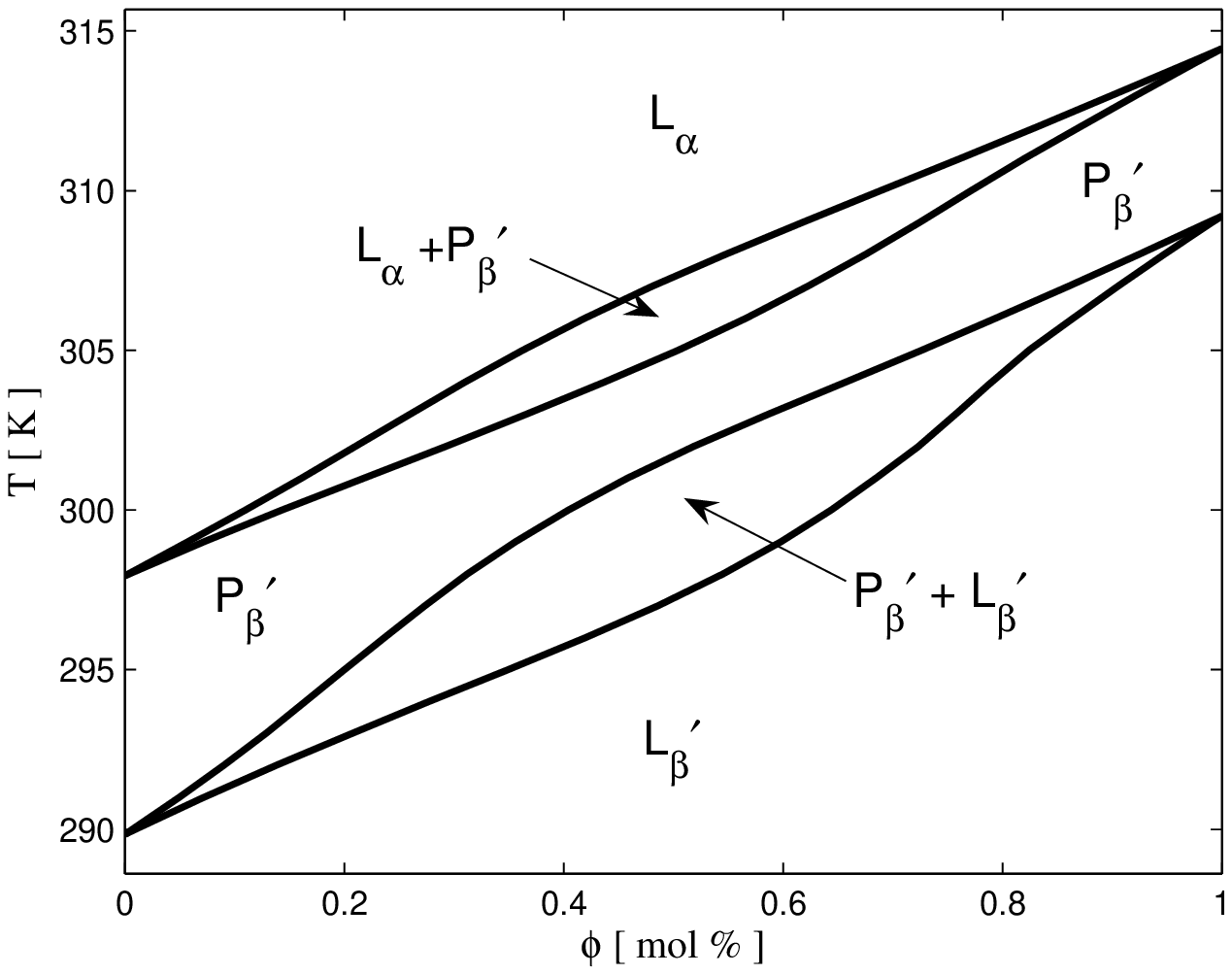}
	\label{fig:TT_J3}
}
\subfigure[]{
\includegraphics[scale=0.55]{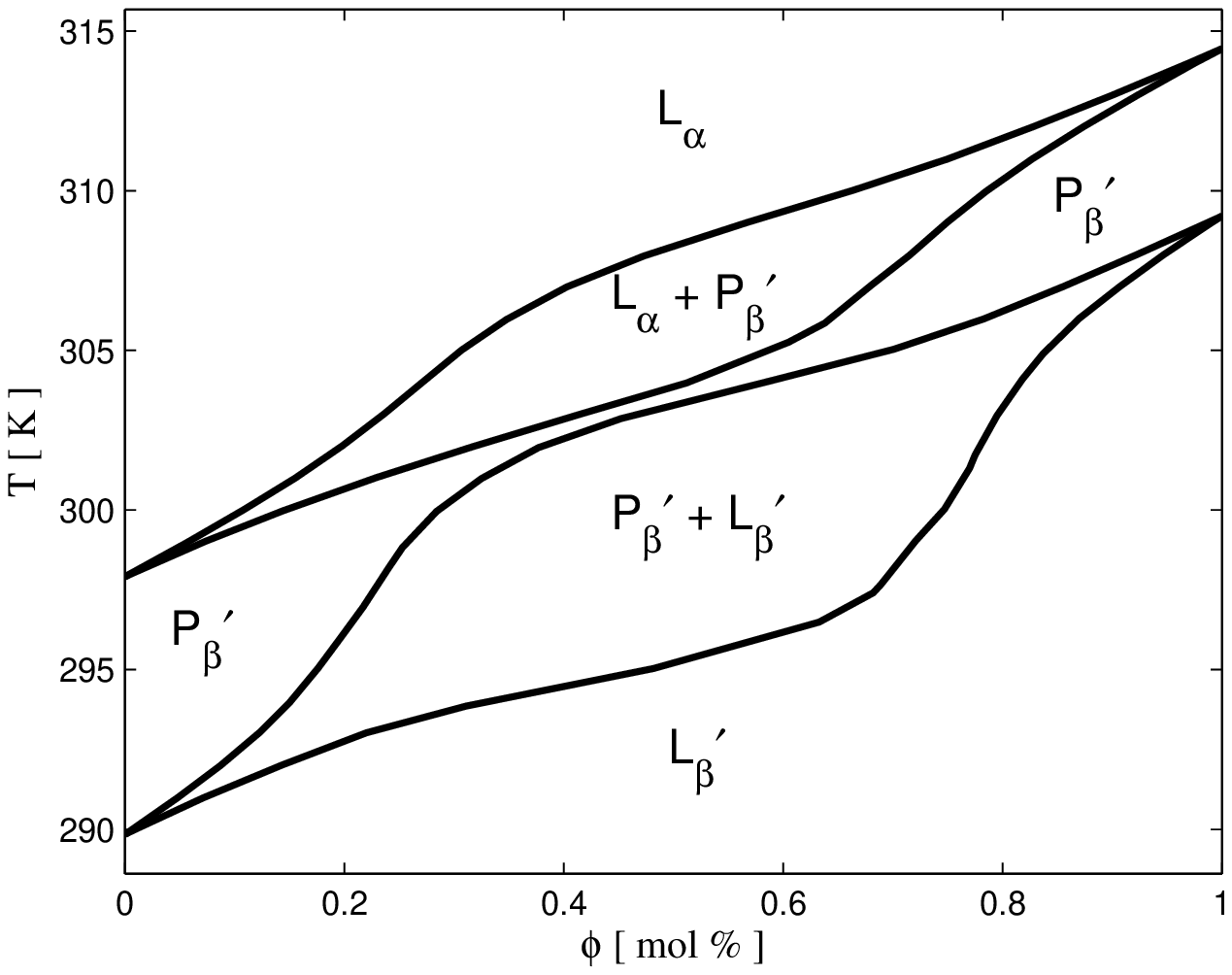}
	\label{fig:TT_J4}
}
\subfigure[]{
\includegraphics[scale=0.55]{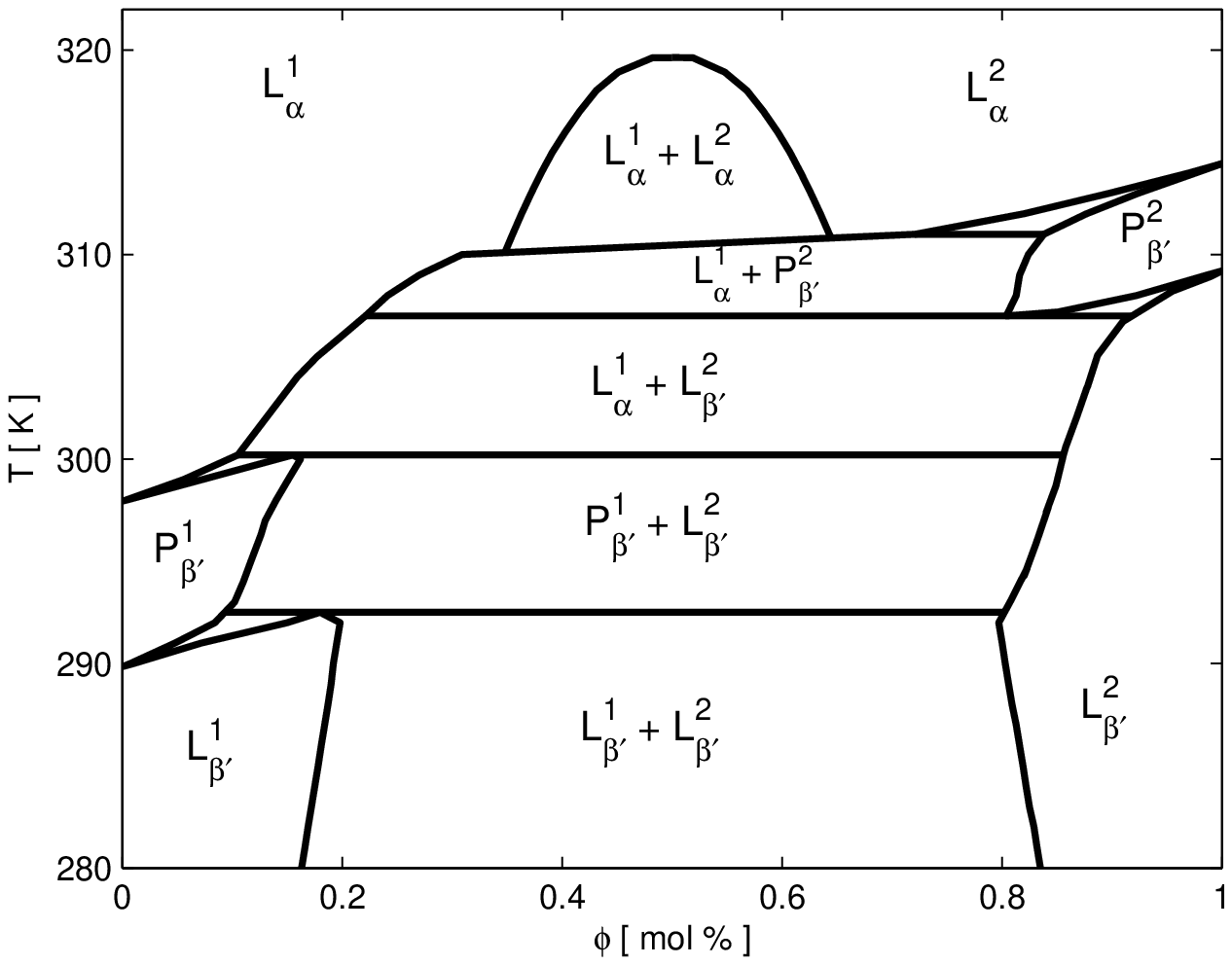}
	\label{fig:TT_J5}
}
\caption{\footnotesize{Calculated mean-field phase diagram of a binary lipid mixture as a function of their relative composition $\phi$ and Temperature $T$. Both the lipids constituting the mixture exhibit a low temperature tilted phase. The lipid interaction parameter is set to be \subref{fig:TT_J3} $J = 3.0\:k_{B} T^{*}$, \subref{fig:TT_J4} $J = 4.0\:k_{B} T^{*}$, \subref{fig:TT_J5} $J = 5.0\:k_{B} T^{*}$. C$_{DMPC}$ = 5.20$\times$10$^{-19}$J and C$_{DPPC}$ = 4.66$\times$10$^{-19}$J. All other parameter values are as in fig.\ref{phasedia}.}}
\end{figure}

In this section, we consider the case where both the constituents exhibit the ripple phase between the $L_{\alpha}$ and $L_{\beta^{\prime}}$ phases. This would correspond to mixtures of lipids such as DPPC and DMPC. We, therefore, choose the value of $C$ for the two lipids to be larger than $C_{Lp}$ and the sign of $\Gamma_4$ to be negative for both of them. The reference temperature for A and B is chosen to be, $T_{A}^{*}= -35^{\circ}\:C$ and $T_{B}^{*}= -17^{\circ}\:C$, respectively. The value of $\Gamma_4$ is taken to be the same for both the lipids. Using these values of the parameters the binary phase diagram was calculated, and the results are shown in fig.\ref{fig:TT_J3}, fig.\ref{fig:TT_J4} and fig.\ref{fig:TT_J5} for different values of $J$.\par
When $J = 3.0 k_{B} T^{*}$ (fig.\ref{fig:TT_J3}) the three phases extend throughout the entire range of composition. These regions are separated by two coexistence regions of $L_{\alpha}$ - $P_{\beta^{\prime}}$ and $P_{\beta^{\prime}}$ - $L_{\beta^{\prime}}$ phases. As $J$ is increases to $4.0\:k_{B} T^{*}$ (fig.\ref{fig:TT_J4}), these coexistence regions almost touch each other, confining the $P_{\beta^{\prime}}$ phase to a narrow strip. With further increase in the value of J (= 5.0$\:k_{B} T^{*}$) two-phase coexistence regions dominate the phase diagram with the pure phases confined to compositions close to the two pure lipids. For this value of $J$ the two lipids show a miscibility gap in the $L_{\alpha}$ phase, with the appearance of a critical point. As discussed in the next section, the calculated phase behavior at low values of $J$ closely resembles that observed in mixtures of very similar lipids. Although the complex phase diagrams calculated for high values of $J$ have not yet been reported in experiments, we believe that such phase behavior should be seen in binary mixtures of very dissimilar lipids, such as those with hydrocarbon and fluorocarbon chains. 
\section{Phase behavior of DPPC-DMPC mixtures}
Small angle scattering techniques can in principle detect microscopic phase separation in the plane of bilayers, if there is sufficient contrast in the scattering densities of the two phases. However, even in the absence of such contrast, macroscopic phase separation can easily be detected from non overlapping reflections in the diffraction pattern coming from the individual  phases. On the basis of the diffraction patterns we have determined  the partial phase diagram of the binary mixture consisting of DPPC  and DMPC. The use of oriented samples has an advantage that the in-plane ordering of molecules can be easily inferred from the wide angle reflections. For example, in the gel phase of both DPPC and DMPC, one observes two wide angle reflections, one on-axis ($q_z=0$) and the other off-axis ($q_z \neq 0$), coming from the quasi hexagonal lattice of hydrocarbon chains of the lipid molecules. The positions of these peaks indicate that the molecules are  tilted with respect to the bilayer normal and that the direction of the tilt is toward nearest neighbor~\cite{Hentschel1983}.\par
\subsection{Materials and Methods}
1,2-dipalmitoyl-sn-glycero-3-phosphocholine (DPPC) and 1,2-dimyristoyl-sn-glycero-3-phosphocholine (DMPC) were purchased from Avanti Polar, and were used as received. Aligned multibilayers of binary lipid mixtures were prepared as follows: A concentrated solution of DPPC-DMPC binary mixtures (dissolved in a mixture of chloroform and methanol (1:1)) in designated molar ratios was deposited on the outer surface of a clean cylindrical glass substrate (radius of curvature of ~9 mm). After deposition, the samples were placed overnight under vacuum to remove any remaining traces of the solvent. Subsequently, they were kept in a water-saturated atmosphere and were hydrated for a couple of days to obtain a stack of bilayers oriented parallel to the surface.\par
Cu $K_{\alpha}$ ($\lambda =$ 1.54 \AA) radiation from a rotating anode x-ray generator (Rigaku, Ultra X18) operating at 48 kV and 70 mA and rendered monochromatic by a multilayer mirror (Xenocs) was used to illuminate the hydrated sample kept inside a sealed chamber with two mylar windows. The chamber temperature was controlled using a circulating water bath to an accuracy of $\pm$ 0.1 $^\circ$C and the relative humidity (RH) inside it was maintained at $98\pm 2\%$, by keeping a reservoir of water. The axis of the cylindrical substrate was oriented perpendicular to the incoming x-ray beam, such that the beam is incident tangentially to the sample.\par
Diffraction patterns were recorded on a 2D image plate detector of 235 mm diameter and 0.1 mm pixel size (Marresearch). All the samples were first heated to a temperature above the main transition temperature of DPPC and the diffraction patterns were recorded during cooling from the L$_\alpha$ phase. The sample temperature and the RH close to the sample were measured with a thermo-hygrometer (Testo 610) inserted into the chamber.
\subsection{Results and discussion}
\begin{figure}[t]
\begin{center}
\includegraphics[width=50mm,angle=90] {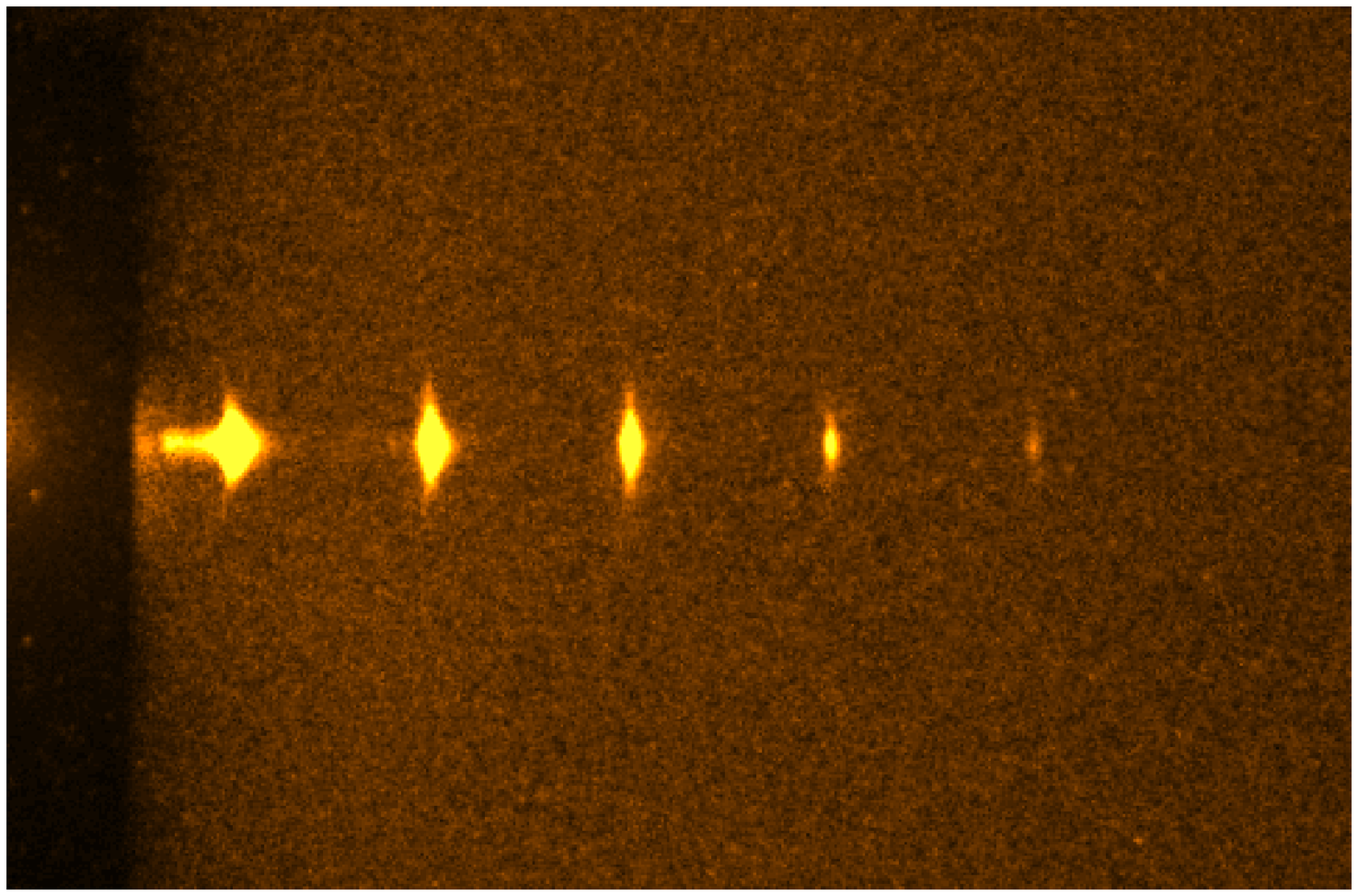}
\includegraphics[width=50mm,angle=90] {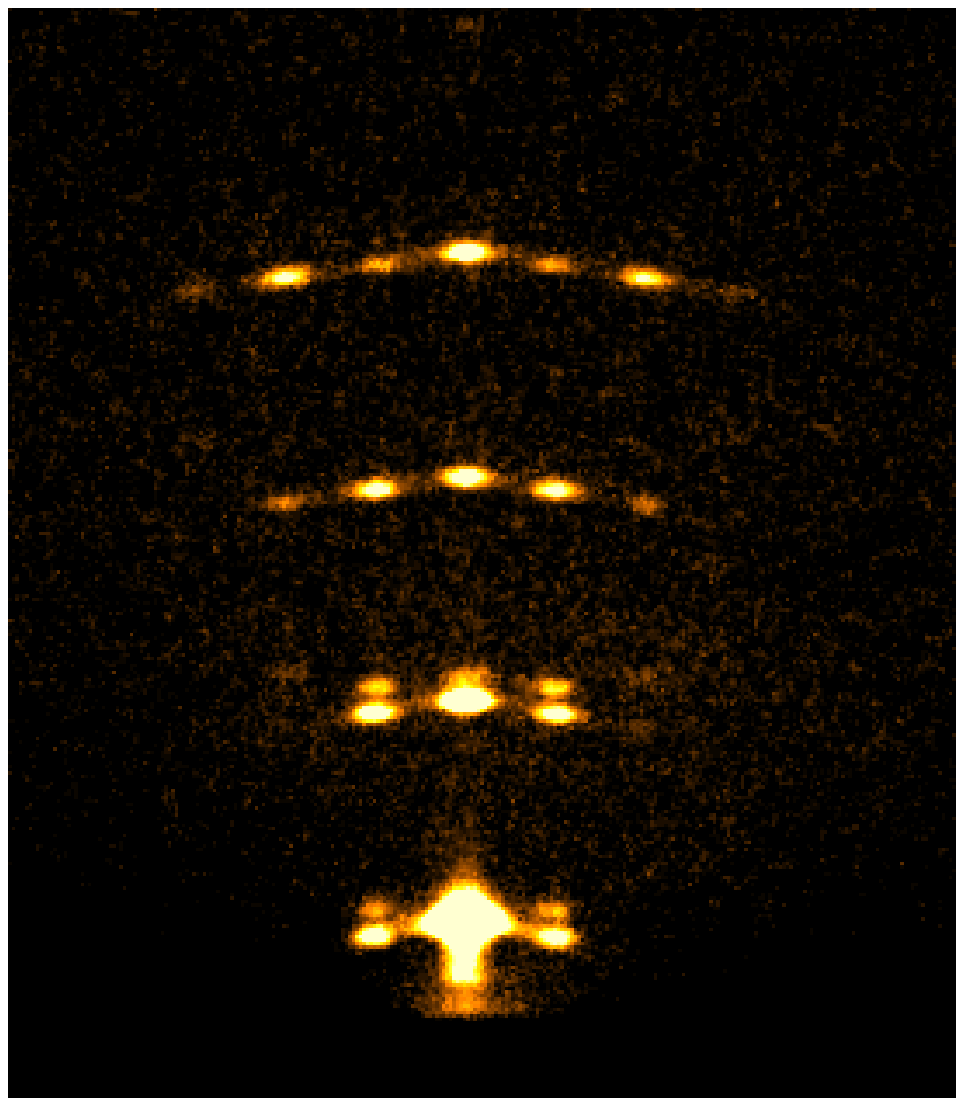}
\end{center}
\caption{\footnotesize{Small-angle diffraction patterns of aligned samples of DPPC - DMPC mixtures in the $L_\alpha$ phase at $T=45.0^\circ$C, $RH=99.9\%$, $\phi=0.5$ (left), $P_{\beta^\prime}$ phase at $T=30.1^\circ$C, $RH=99.9\%$, $\phi=0.5$ (right). The bilayer normal is vertical.}}
\label{diff-patts-powder}
\end{figure}
\begin{figure}[t]
\begin{center}
\includegraphics[width=40mm,angle=90] {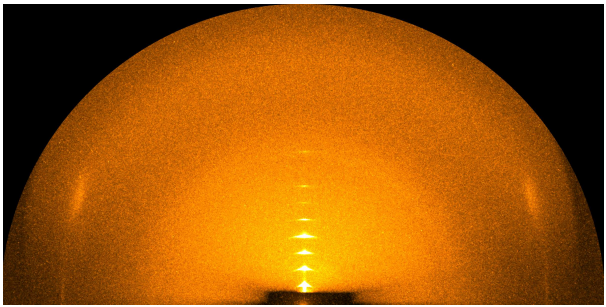}
\end{center}
\caption{\footnotesize{Diffraction pattern of an aligned sample of DPPC - DMPC mixtures in the $L_{\beta^\prime}$ phases, showing both the small-angle lamellar peaks due to the periodic stacking of the bilayers and the wide-angle peaks due to the in-plane ordering of the tilted chains in the bilayer. $T=14.9^\circ$C and $RH = 99.9\%$, $\phi=0.5$ . The bilayer normal is vertical.}}
\label{diff-patts-aligned}
\end{figure}
\begin{figure}[t]
  \centering    
		\includegraphics[width=80mm]{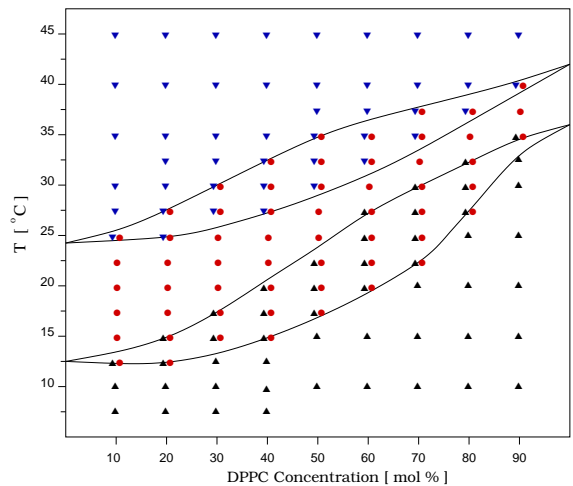}
		\caption{\footnotesize{Phase diagram of DPPC-DMPC bilayers at 98\% RH determined from x-ray diffraction data. $\bigtriangledown$, $\bullet$ and $\bigtriangleup$ correspond to the $L_\alpha$, $P_{\beta^\prime}$ and $L_{\beta^\prime}$ phases, respectively.}}
	\label{fig:LA_PB}
\end{figure}
Figure[\ref{fig:LA_PB}] shows the partial phase diagram of DPPC - DMPC mixtures at $98\pm 2\%$ Rh. The three pure phases extend over the complete composition range, with two two-phase regions separating them. The phases were identified from their diffraction patterns. The $L_{\alpha}$ phase gives a few reflections in the small angle regions, which correspond to different diffraction orders from a periodic lamellar stack. The lamellar periodicity in this phase is $\sim$ 5.6 nm for DPPC and $\sim$ 5.0 nm for DMPC; the mixtures have intermediate values depending on the composition. The wide angle reflection due to chain ordering is very diffuse in this phase and cannot be seen in the present geometry for exposure times of the order of 10 min. Additional satellite reflections appear in the $P_{\beta^{\prime}}$ phase due to the periodic height modulations of the bilayer. These reflections can be indexed on a two-dimensional oblique lattice. The wavelength of the bilayer height modulations in this phase is typically $\sim$ 150 nm. The lower temperature $L_{\beta^{\prime}}$  phase shows more lamellar reflections in the small angle region compared to the $L_{\alpha}$ phase due to decreased thermal undulations of the rigid bilayers. In addition sharp wide angle reflections are observed in this phase due to the ordering of the chains in the plane of the bilayer. The presence of a on-axis ($q_z=0$) and an off-axis ($q_z \ne 0$) reflection show that the chains are tilted , with the tilt direction along the nearest neighbour. The DPPC-DMPC phase diagram closely resembles that calculated from our model for low values of $J$. The low value of $J$ in this case is understandable since the two lipids are very similar with the only difference being that each chain of DPPC is longer by two $CH_3$ groups. 



\section{Conclusion}
We have developed a continuum theory of phase transitions in single- and two-component achiral lipid bilayers. This model is found to reproduce all the salient structural features of the ripple phase occurring between the fluid $L_{\alpha}$ phase and the gel $L_{\beta^{\prime}}$ phase. In addition, it predicts the existence of a novel variant of this phase, which has not been experimentally observed so far. This theory has been used to calculate some generic phase diagrams of binary phospholipid mixtures. We have also determined the phase diagram of DPPC-DMPC mixtures from x-ray diffraction studies on aligned multilayers. It is found to be in good agreement with that calculated from the model for weakly segregating species.



\end{document}